\documentclass{article}
\usepackage{emulateapj,amsfonts,psfig}
\begin{document}
\newcommand{\be}{\begin{equation}}
\newcommand{\en}{\end{equation}}
\def\ltsima{$\; \buildrel < \over \sim \;$}
\def\lsim{\lower.5ex\hbox{\ltsima}}
\def\loe{\lower.5ex\hbox{\ltsima}}
\def\gtsima{$\; \buildrel > \over \sim \;$}
\def\gsim{\lower.5ex\hbox{\gtsima}}
\def\goe{\lower.5ex\hbox{\gtsima}}
\def\rref{\par\noindent\hangindent=1.5truecm}
\def\aa #1 #2 {A\&A #1 #2}
\def\aass #1 #2 {A\&AS #1 #2}
\def\araa #1 #2 {ARA\&A #1 #2}
\def\mon #1 #2 {MNRAS #1 #2}
\def\apj #1 #2 {ApJ #1 #2}
\def\apjss #1 #2 {ApJS #1 #2}
\def\apjl #1 #2 {ApJ #1 #2}
\def\astrj #1 #2 {AsJ #1 #2}
\def\nat #1 #2 {Nature #1 #2}
\def\pasj #1 #2 {PASJ #1 #2}
\def\pasp #1 #2 {PASP #1 #2}
\def\msai #1 #2 {Mem. Soc. Astron. Ital. #1 #2}
\def\ass #1 #2 {Ap. Sp. Science #1 #2}
\def\sci #1 #2 {Science #1 #2}
\def\phrevl #1 #2 {Phys. Rev. Lett. #1 #2}
\newcommand{\si}{\left(\frac{\sigma}{0.005}\right)}
\newcommand{\psrdot}{\frac{\dot{P}_{-20}}{P_{-3}^3}}
\newcommand{\ggg}{$\gamma$}
\newcommand{\eee}{$e^{\pm}$}
\newcommand{\ergs}{\rm \ erg \; s^{-1}}
\newcommand{\msol}{\su M_{\odot} }
\newcommand{\etal}{et al.\ }
\newcommand{\Po}{$ P_{orb} \su$}
\newcommand{\pot}{$ \dot{P}_{orb} / P_{orb} \su $}
\newcommand{\myr}{ \su M_{\odot} \su \rm yr^{-1}}
\newcommand{\ppp}{ \dot{P}_{-20} }
\newcommand{\ci}[1]{\cite{#1}}
\newcommand{\bb}[1]{\bibitem{#1}}
\newcommand{\pdot}{ $\dot{P}_{orb}$ \su}
\newcommand{\befl}{ \vspace*{-17pt} \begin{flushright}}
\newcommand{\enfl}{\end{flushright}}
\def\deg {^\circ}
\def\mdot {\dot M}
\def\kms  {\rm \ km \, s^{-1}}
\def\cms  {\rm \ cm \, s^{-1}}
\def\gs   {\rm \ g  \, s^{-1}}
\def\cmtre {\rm \ cm^{-3}}
\def\cmdue {\rm \ cm^{-2}}
\def\gcmdue {\rm \ g \, cm^{-2}}
\def\gcm  {\rm \ g \, cm^{-3}}
\def\rsole {~R_{\odot}}
\def\msole {~M_{\odot}}
\def\fH {{\cal H}}
\def\op {{\cal K}}
\def\nupa{\vfill\eject\noindent}
\def\der#1#2{{d #1 \over d #2}}
\def\inizio{\2acapo\penalty+10000}
\def\fine{\acapo\penalty-10000\blank}
\received{~~} \accepted{~~} 
\journalid{}{}
\articleid{}{}

\title{Kilohertz quasi-periodic oscillations in low mass X--ray binary sources 
and their relation with the neutron star magnetic field}

\author{Sergio Campana}

\affil{Osservatorio Astronomico di Brera, Via Bianchi 46, I-23807 
Merate (LC), Italy}

\begin{abstract}
Starting from the observation that kilohertz Quasi Period Oscillations (kHz QPOs)
occur in a very narrow range of X--ray luminosities in neutron star low
mass X--ray binaries, we try to link the kHz QPO observability to variations of 
the neutron star magnetospheric radius, in response to changing mass inflow rate.
At low luminosities, the drop off of kHz QPOs activity may be explained by the 
onset of the centrifugal barrier, when the magnetospheric radius reaches 
the corotation radius. At the opposite side, at higher luminosities, 
the magnetospheric radius may reach the neutron star and 
the vanishing of the magnetosphere may led to the stopping of the 
kHz QPOs activity. If we apply these constraints, the
magnetic fields of atoll ($B\sim 0.3-1\times 10^8$ G for Aql X-1) and 
Z ($B\sim 1-8\times 10^8$ G for Cyg X-2) sources can be derived. 
These limits naturally apply in the framework of beat frequency models but 
can also work in the case of general relativistic models.
\end{abstract}

  
\section {Introduction}

Color-color diagrams (CDs) have been widely used to gain insight on low
mass X--ray binaries (LMXRBs) hosting a neutron star 
(e.g. Hasinger \& van der Klis 1989).
These sources are usually classified as atoll or Z sources depending on
the path that they trace in the CDs as their luminosity vary. 
Atoll sources are characterised by an upward curved branch (lower and upper
banana state) and one or more harder island states
(cf. Fig. 1). Z sources produce a Z-shaped path
in the CD with three branches, from hard to soft: horizontal
branch (HB), normal branch (NB) and flaring branch (FB).
Type I X--ray burst properties correlate well with position in the CD 
(van der Klis et al. 1990). This led to argue
that position on the CD along the atoll was a good
indicator of $\mdot$ (but X--ray flux was not).
For Z sources (a subset of which show sporadic X--ray bursts too), 
optical and UV emission
as well as quasi periodic oscillations (QPOs) models indicate that
$\mdot$ increases as HB$\to$NB$\to$FB (cf. van der Klis 1995).

RossiXTE observations have shed new light on these sources, revealing
QPOs in the kHz frequency range (for a review see van der Klis 2000). In
atoll sources, kHz QPOs occur in the island  
and lower banana part of the CD (M\'endez 1998 and references therein). 
In the case of Z sources kHz QPOs are present all the way down the HB and 
disappear in the lower part of the NB (e.g. Wijnands \& van der Klis 1998b
and references therein). 
KHz QPOs are therefore detected in a narrow range of 
fluxes and in similar positions in the CDs. Whereas the disappearance of the 
kHz QPO activity at the low luminosity end may be related in some atoll sources 
to a lack of sensitivity, this is certainly not the case at the bright end.

In this letter we try to link the source behavior in the CD to the
physics of the accretion process. We mainly concentrate on atoll sources,
since for Z sources data are more sparse and they still lack an indication 
of the spin period from e.g. type I X--ray bursts.

\section{Accretion regimes}

Accretion phenomena onto a fast spinning, weakly magnetic neutron star 
three radii which do not vary with the mass inflow rate can be identified:

\begin{itemize}

\item{the neutron star radius, $R$, which depends on the neutron star
mass and equation of state. We scale here $R=10\,R_6$ km for a 
neutron star of mass $M=2\,M_2\msole$;
}

\item{
the marginally stable orbit radius, $r_{\rm ms}$. 
In the case of a rapidly rotating neutron star 
a small correction to the Schwarzschild value has to be included. As a first 
order in the specific angular momentum $j$ 
[$j=2\,\pi\,c\,I/(G\,M^2\,P)$, with $c$ the speed of light, $I$ and $P$ the
neutron star moment of inertia and spin period and $G$ the gravitational 
constant], $r_{\rm ms}\simeq (6-2.31\,j)\,G\,M/c^2$ (Klu\'zniak 1998).
In the case of `soft' equations of state this radius is larger 
than $R$, whereas for stiff equation of states it is smaller and does 
not play any role.
In the case of a spin frequency of 1.8 ms (as in 
the case of Aql X-1; Zhang et al. 1998a), and 
for a $2\msole$ neutron star with $I=2\times10^{45}$ g cm$^{-2}$,
$r_{\rm ms}\sim 16\,M_2$ km;
}

\begin{figure*}[!htb]
\centerline{\psfig{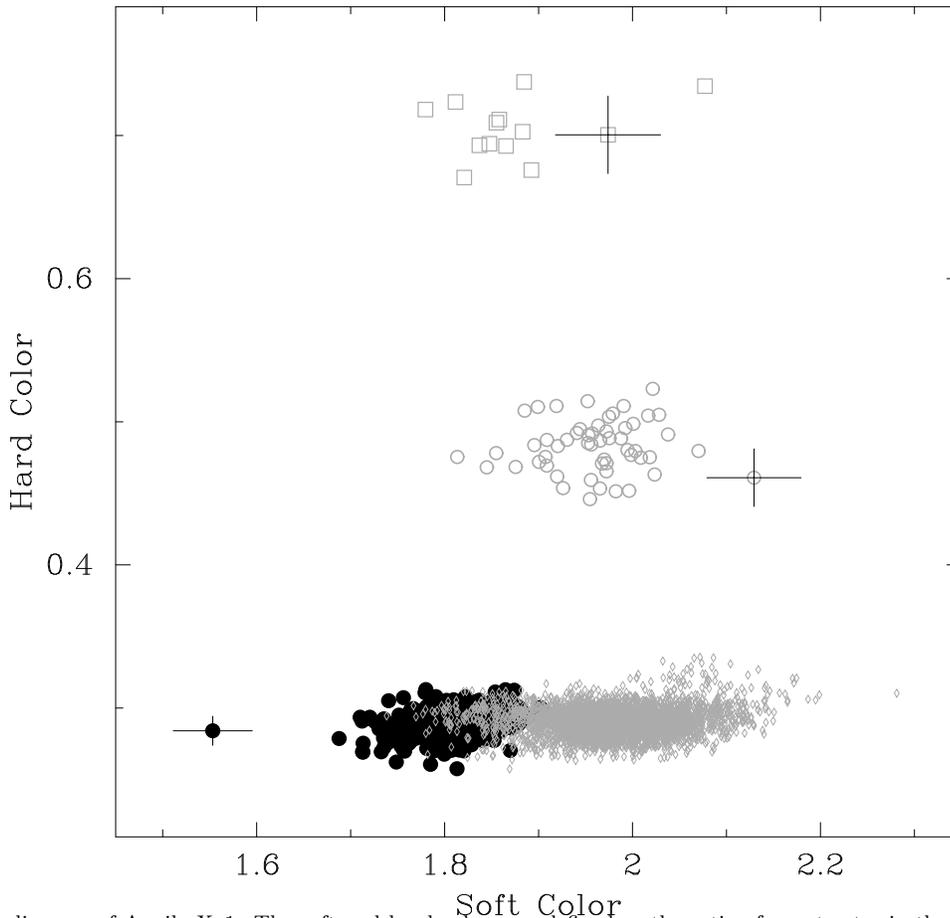}}
\caption{Color-color diagram of Aquila X--1.  The soft and hard colors are
defined as the ratio of count rates in the bands $3.5 - 6.0$ keV and
$2.0 - 3.5$ keV, and $9.7 - 16.0$ keV and $6.0 - 9.7$ keV, respectively.
Points in the banana branch are marked with closed circles,
in the lowest island state with filled squares, in the middle island with open 
circles and in the extreme island with open squares.  Typical error bars
in the banana and the island states are shown.  Black and grey symbols indicate
segments with and without kHz QPOs, respectively. Taken from Reig et al. (2000).
}
\label{uno}
\end{figure*}

\item{the corotation radius, $r_{\rm cor}$, at which a test particle in 
Keplerian orbit corotates with the neutron star, 
$r_{\rm cor}=\Bigr({{G\,M\,P^2}\over {4\,\pi^2}}\Bigl)^{1/3}$. 
In the case of a spin frequency of 1.8 ms ($P=1.8\,P_{\rm 1.8 ms}$ ms)
we have $r_{\rm cor}=28\,M_2^{1/3}\,P_{\rm 1.8 ms}^{2/3}$ km.}

\end{itemize}

As the mass inflow rate changes
the limiting radius at which the neutron star magnetic field pressure 
balances the ram pressure of the mass inflow towards the neutron star 
(i.e. the magnetospheric radius, $r_{\rm m}$) changes too.
In the case of spherical accretion with a dipolar magnetic field
$r_{\rm m}\sim\Bigr({{\mu^4}\over{2\,G\,M\,\mdot^2}}\Bigl)^{1/7}$
(where $\mdot$ is the mass inflow rate onto the neutron star magnetosphere 
and $\mu=B\,R^3$ is the magnetic moment). 
A parameter $\xi$ is usually 
considered to account for the case of disk accretion, with $\xi\sim 0.5-1$ 
(Ghosh \& Lamb 1979, 1992). Wang (1995) pointed out that $\xi$ is
in the range 0.87--0.95. In the following, we assume $\xi=1$.
For an accretion luminosity of $L=G\,M\,\mdot/R=10^{37}\,L_{37}\ergs$, 
we derive a magnetospheric radius of 
$r_{\rm m}=20\,B_8^{4/7}\,R_6^{10/7}\,M_2^{1/7}\,L_{37}^{-2/7}\,\xi$ km 
(where $B=B_8\,10^8$ G is the magnetic field). 
For higher luminosities the radiation pressure starts playing a role (region A
in the standard disk model) and the magnetospheric radius changes to
$r_{\rm m}^{\rm A}=33\,B_9^{20/39}\,L_{38}^{-2/13}\,M_{2}^{1/39}
\,R_6^{18/13}\,\xi$ km
(cf. Ghosh \& Lamb 1992; here we neglect the weak dependence on the 
viscosity parameter).

\section{Application to Atoll sources}

As a working example we can consider the case of the transient atoll
source Aql X-1. This is the
prototype of soft X--ray transient sources (Campana et al. 1998a) and,
together with the accreting X--ray binary pulsar SAX J1808.4--3658 
(Wijnands \& van der Klis 1998a; Chakrabarty \& Morgan 1998), is the only 
LMXRB which showed signs of centrifugal inhibition of accretion 
(Campana et al. 1998b;  Zhang, Yu \& Zhang 1998; Gilfanov et al. 1998). 
During a type I burst of Aql X-1 observed with RossiXTE, a 
coherent periodicity was discovered at 549 Hz (Zhang et al. 1998a). 
This frequency may correspond to the spin frequency of the neutron star
or, like the case of 4U 1636--53, to twice its value (Miller 1999). In the 
following we adopt a spin period of 1.8 ms.

In Fig. 1 we show the CDs of Aql X-1 based on RossiXTE.
Data fall into three main distinct groups, which can 
be readily identified with the extreme island, island and banana states 
based on the hard color.
By adopting an absorbed blackbody plus power law model,
Reig et al. (2000) derived the 2--10 keV X--ray luminosity
in the different states of the CD diagram (for a distance of 2.5 kpc).
This varied from $2.0\times10^{34}$ erg s$^{-1}$ in
the extreme island state (not shown in Fig. 1) to $3.2\times10^{35}$ erg 
s$^{-1}$ in the island state.
In the banana state, the luminosity increased from left to 
right, from $2.4\times10^{36}$ erg s$^{-1}$ to $3.9\times 10^{36}$ 
erg s$^{-1}$ (Reig et al. 2000). Assuming the spectrum at a comparable 
luminosity observed with BeppoSAX (Campana et al. 1998b), we derive a 
bolometric correction of $\sim 4$ in the 0.1--40 keV energy band. 
We remark, however, that it has long been recognised that the observed flux does 
not provide a confident measure of the accretion luminosity. Different contributions
may come from e.g. the neutron star rotational energy which is extracted by 
the disk (Priedhorsky 1986). 

Despite the large variations in X--ray luminosity, kHz QPOs are observed in 
selected regions of the CDs. In the transient sources Aql X-1 and
4U 1608--52 (M\'endez et al. 1999) there exist RossiXTE observations in which 
kHz QPO activity was not detected. The corresponding limiting RossiXTE PCA rates for 
the two sources are $\sim 600$ and $\sim 400$ c s$^{-1}$, respectively
(2--16 keV; M\'endez 1998). Moreover, 
the persistent atoll source 4U 0641+091 showed kHz QPOs in the range 250--1800 
c s$^{-1}$ (2--60 keV; van Straaten et al. 2000). We conclude that, at least
in the case of Aql X-1, the disappearance of the kHz QPO activity at the faint end is 
likely not due to poor statistics. For what concerns the maximum flux at which kHz
QPOs have been detected, no problems of statistics exists ($\sim 1500$ and 
$\sim 3000$ c s$^{-1}$ for Aql X-1 and 4U 1608--52, respectively, cf. M\'endez 1998).
The stopping occurs for all the atoll sources in the lower banana state.  

The existence of selected regions in the CD (even if slightly different 
from source to source) in which kHz QPOs are detected is noteworthy.
We now try to identify the onset and the disappearance of kHz QPO 
activity with changes in the magnetospheric radius. 

($i$) As the accretion rate increases, the onset of the kHz activity can be 
related to the exit of the source from the propeller regime, when most of the 
matter is inhibited to accrete onto the neutron star surface by centrifugal forces 
(i.e. $r_{\rm m}\sim r_{\rm cor}$) and therefore kHz QPOs can be hardly 
detected.
The luminosity jump expected for the transition from the 
propeller to the accretion regime in the case of a ms spinning neutron star 
is just a factor of a few (Corbet 1996; Campana \& Stella 2000). 
This flux reduction is independent on the mechanism invoked to explain the
kHz QPOs, like beat-frequency models (BFMs; Strohmeyer et al. 1996; Miller, 
Lamb \& Psaltis 1998) or general relativistic models (GRMs, Stella \& Vietri 
1998, 1999).

The minimum luminosity for which kHz QPOs have been detected during the 
fading end of an Aql X-1 outburst is $1.1\times 10^{36}\ergs$
(2--10 keV; Zhang et al. 1998a). This luminosity is a factor of $\sim 2$ higher than 
the luminosity corresponding at the steep increase in the hardness ratio and 
associated to the centrifugal barrier onset as interpreted by Zhang, Yu \& 
Zhang (1998) and Campana et al. (1998b). Considering the bolometric correction 
given above we assume a 0.1--40 keV luminosity of $L_{\rm min}=4.2\times 
10^{36}\ergs$. This luminosity represents an upper limit to the accretion 
luminosity (which defines the magnetospheric radius), but at the onset
of the centrifugal barrier external contribution such as the luminosity
extracted from the neutron star rotational energy should weaken considerably.

Equating the magnetospheric to the corotation radius, we derive 
an upper limit to the magnetic field value of:
\be
B\lsim 1\times 10^8\,L_{\rm min}^{1/2}\,R_6^{-5/2}\,M_2^{1/3}\,
P_{\rm 1.8 ms}^{7/6}\,\xi^{-7/4}\ {\rm G}.
\en 
In the case of Miller's et al. BFM the relevant radius for kHz QPOs is the 
sonic radius, $r_{\rm s}$, which lies inside the magnetospheric radius. Since, 
$r_{\rm s}<r_{\rm m}<r_{\rm cor}$, the constraint above applies as well.

($ii$) The disappearance of the kHz QPOs at high luminosities may be related 
to the disappearance of the magnetosphere, when $r_{\rm m}$ becomes comparable to 
the neutron star radius, $R$. 
A similar line of reasoning has been applied by Cui et al. (1998) interpreting
the transition from island and banana and the disappearance of the kHz QPO, 
as the disk becoming disangaged from the magnetic field.
In particular, by comparing $r_{\rm m}$ with $r_{\rm ms}$
Cui et al. (1998) were able to limit the neutron star magnetic field to $\sim 
0.6-0.9\times 10^8$ G. 

The disappearance of the kHz QPOs can be interpreted in the framework of
BFMs as the drop out of the magnetosphere without which the 
beating mechanism cannot work. In the Miller's et al. BFM the relevant 
radius is the sonic point radius which, by definition, lies always inside the 
magnetospheric radius. On the other hand, the disappearance of the 
magnetosphere may leave no space for an orbiting blob to acquire 
the necessary eccentricity to generate kHz QPOs as predicted by GRMs 
(Stella \& Vietri 1999) or if the magnetosphere itself 
is needed to generate the small eccentricity in a resonant interaction 
(Vietri \& Stella 1998). In general both the sonic point model and the  
GRMs do not require the presence of a magnetosphere to work.
However, we expect that the neutron stars in LMXRBs do have a magnetic field
if they are the progenitors of millisecond radio pulsars. This link has been
recently confirmed by the detection of coherent pulsations in the transient 
LMXRB SAX J1808--3658 (Wijnands \& van der Klis 1998a).

In the case of Aql X-1, kHz QPOs stop to be detected at a luminosity 
higher than $3.6\times10^{36}\ergs$ (2--10 keV; Zhang et al. 1998a). 
Given the small range of luminosity we adopt the same bolometric correction, 
resulting in 0.1--40 keV luminosity of $L_{\rm max}=1.4\times10^{37}\ergs$.
In this case, we are going to derive a lower limit on the magnetic field and 
no problems are posed by taking a higher luminosity than the true 
accretion one.
We can obtain a conservative estimate of the 
magnetic field from $r_{\rm m}\lsim R$:
\be
B\gsim 3\times 10^7\,L_{\rm max}^{1/2}\,R_6^{-3/4}\,M_2^{-1/4}\,\xi^{-7/4}
\ {\rm G}.
\en
Different parameters can play a role in changing the exact value of these
constraints as well as the uncertainties in the theory underlying the location
of the magnetospheric radius. Despite these uncertainties, the two limits are 
well consistent and constrain the magnetic field of Aql X-1 to $B\sim 
0.3-1\times 10^8$ G. 


\subsection{4U 1820--30}

A further case is represented by the atoll source 4U 1820--30 for which 
indication of a saturation in the kHz QPO frequencies versus the flux 
intensity has been observed (Zhang et al. 1998b; Kaaret et al. 1999 and 
references therein). 
4U 1820--30 follows the general behaviour of atoll sources described
above (Zhang et al. 1998b). 
The saturation starts near the lower banana onset (Zhang et al. 1998b).
KHz QPOs appear at a 0.3--40 keV luminosity of $5\times10^{37}\ergs$ 
(for a distance of 8 kpc) and disappear at $8\times10^{37}\ergs$, i.e.
spanning a vary small range of luminosities. We adopt these as 
bolometric luminosities and derive, as described above, 
$B\lsim 9\times 10^8$ G (for a neutron star spin period of 3.6 ms, as
inferred from the kHz QPO separation) and 
$B\gsim 0.8\times 10^8$ G, respectively. 
A further constraint comes from the interpretation of the saturation of the 
kHz QPO frequencies as due to the magnetospheric radius reaching the 
last stable orbit. If this is the case, the upper kHz QPO 
frequency is related to the Keplerian frequency at the disk inner edge.
In the case of GRMs, the ceasing of the kHz QPOs derives 
from the lack of space between the point at which the blob leaves the accretion 
disk and the neutron star.

A saturation value of 1065 Hz has been measured in 4U 1820--30, 
leading to an estimate of the neutron star mass of $\sim 2.1\msole$.
Alternatively, this implies a radius of the marginally stable orbit
of $r_{\rm ms}=17.4$ km. 
The saturation occurs at a luminosity of $6.6\times10^{37}\ergs$ (0.3--40
keV), which provides a direct estimate of the magnetic field value of
$B\sim 2\times 10^8$ G (close to the lower limit derived above as expected 
due to the fact that $r_{\rm ms}\sim R$). In the case of Miller's et al. 
BFM the estimate of the magnetic field gives only a lower limit, since the
sonic radius is smaller than the magnetospheric radius. 

The saturation in the kHz QPO frequencies has been observed only in 
4U 1820--30. A possible explanation relies on the large neutron star
mass inferred for this object, which in turn implies a smaller
radius than for a $\sim 1.6-1.8\msole$ neutron star by a factor of 10--20\%. 
This might be enough to prevent the observation of the kHz QPO saturation
in lighter neutron stars.

\section {Application to Z sources}

In Z sources kHz QPOs are usually observed in the upper part of the NB
down to the end of the HB (e.g. Cyg X-2, Wijnands et al. 
1998; GX 17+2, Wijnands et al. 1997). Flux determinations are very rare.
Only in the case of GX 349+2 (Zhang, Strohmeyer \& Swank 1997) the 
onset of kHz QPO occurs at a 2--10 keV luminosity
of $\sim 1.2\times 10^{38}\ergs$ (for a distance of 8.5 kpc).
This is consistent with the idea that the lower apex of the Z path
occurs when the source attains the Eddington luminosity.
Despite the variations observed in the CDs, Z sources vary very 
little in luminosity, with typical RossiXTE PCA rates in excess of 1000 c 
s$^{-1}$ (Wijnands \& van der Klis 1998b)
such that problems with a too poor statistics do not exist.

We consider the case of Cyg X-2, which showed kHz QPO down to apex NB/FB.
The 0.1--20 keV luminosity extrapolated by EXOSAT 
observations at the NB/FB apex is $L_{\rm max}\sim 2\times 10^{38}\ergs$ 
(at 8 kpc; Chiappetti et al. 1990). The luminosity is high and the inner disk
regions are likely in the radiation pressure regime such that the magnetospheric 
radius $r^A_{\rm m}$ should be adopted.
Using this luminosity to infer the lower bound on the magnetic field we obtain 
$B\gsim 1\times 10^8$ G. This is a conservative lower limit 
since at these luminosities the coupling between disk and neutron star should 
be high. KHz QPOs have been observed at the end of the HB, for which the minimum 
luminosity observed with EXOSAT is just a factor of 2 smaller. From this we derive 
$B\lsim 8\times 10^8$ G, assuming a neutron star spin 
period of 2 ms, still waiting for the discovery of a coherent periodicity 
from bursting Z sources. Higher fields may be obtained by assuming 
faster spin periods.

\section{Discussion}

We interpret the onset and the end of kHz QPOs in terms of the variable 
extent of the neutron star magnetosphere. 
The turn-on of kHz QPOs at low fluxes is interpreted 
as the starting of accretion onto the neutron star surface, overcoming 
the effects of the centrifugal barrier (here we do not claim that this
barrier is working perfectly, but if it can stop the large fraction
of the infalling matter, the flux reduction will prevent to detect
kHz QPOs). 
The disappearance of the kHz QPOs corresponds instead to the
`disruption' of the magnetosphere on the neutron star (see also Cui et al. 1998). 
Given these assumptions, we estimate in the case of the atoll source
Aql X-1 a magnetic field of $B\sim 0.3-1\times 10^8$ G.
Similar results can be obtained for other atoll sources. 
Interpreting the saturation of the kHz QPO frequency for increasing
luminosities in 4U 1820--30, leads to a magnetic field estimate
of $\sim 2\times 10^8$. 
In the case of Z sources we derive a rough estimate based on luminosity
obtained with EXOSAT. In the case of Cyg X-2, we obtain a very conservative 
lower limit of $B\sim 1-8\times 10^8$ G. 
This is consistent with the idea that Z sources have higher magnetic 
field than atoll sources.

%

This scenario allows for clear predictions 
on the luminosities at which the kHz QPOs set in and out, which can be 
verified with RossiXTE monitoring of low mass X--ray binaries.

\begin{acknowledgments}
I thank P. Reig for permission of publishing Fig. 1. I acknowledge
useful discussions with T. Belloni, M. M\'endez \& L. Stella.
\end{acknowledgments}

\end{document}